\documentclass[conference]{IEEEtran}
\IEEEoverridecommandlockouts
\usepackage{cite}
\usepackage{amsmath,amssymb,amsfonts}
\usepackage{algorithmic}
\usepackage{graphicx}
\usepackage{textcomp} 
\usepackage{enumitem}
\usepackage{xcolor}
\usepackage{url}
\usepackage{authblk}
\usepackage{subcaption} 
\usepackage[hidelinks]{hyperref}
\usepackage{flushend}
\usepackage[all=normal,paragraphs=tight,floats=normal,mathspacing=normal,wordspacing=tight,charwidths=tight,mathdisplays=normal,leading=tight]{savetrees}

\def\BibTeX{{\rm B\kern-.05em{\sc i\kern-.025em b}\kern-.08em
    T\kern-.1667em\lower.7ex\hbox{E}\kern-.125emX}}
\begin{document}

\title{Model-based learning for joint channel estimation\\ and hybrid MIMO precoding}

\author[1]{Nay Klaimi}
\author[1]{Amira Bedoui}
\author[2]{Clément Elvira}
\author[1]{Philippe Mary}
\author[1]{Luc Le Magoarou\thanks{This work is supported by the French national research agency (MoBAIWL project, grant ANR-23-CE25-0013)}}

\affil[1]{Univ Rennes, INSA Rennes, CNRS, IETR-UMR 6164, Rennes, France}
\affil[2]{IETR UMR CNRS 6164, CentraleSupelec Rennes Campus, 35576 Cesson Sévigné, France}

\newcommand{\remCE}[1]{{\scriptsize \color{blue} [CE: #1]}}
\newcommand{\addCE}[1]{{\color{blue}#1}}

\maketitle

\begin{abstract}

Hybrid precoding is a key ingredient of cost-effective massive multiple-input multiple-output transceivers. However, setting jointly digital and analog precoders to optimally serve multiple users is a difficult optimization problem. Moreover, it relies heavily on precise knowledge of the channels, which is difficult to obtain, especially when considering realistic systems comprising hardware impairments.
In this paper, a joint channel estimation and hybrid precoding method is proposed, which consists in an end-to-end architecture taking received pilots as inputs and outputting precoders. The resulting neural network is fully model-based, making it lightweight and interpretable with very few learnable parameters. The channel estimation step is performed using the unfolded matching pursuit algorithm, accounting for imperfect knowledge of the antenna system, while the precoding step is done via unfolded projected gradient ascent. The great potential of the proposed method is empirically demonstrated on realistic synthetic channels. 

\end{abstract}


\section{Introduction}
Massive Multiple-Input Multiple-Output (MIMO) technology holds significant promise in enhancing the efficiency of wireless communication systems \cite{larsson2014}, due to its ability to support a large number of users simultaneously while improving spectral efficiency and reducing interference. However, the implementation of Massive MIMO can be costly due to the need for a large number of radio frequency (RF) chains. Hybrid precoding \cite{molisch2017} emerges as a cost-efficient solution to this challenge. By combining analog and digital precoding techniques, hybrid precoding reduces the number of required RF chains, making Massive MIMO more feasible for practical deployment.

Despite these advantages, determining appropriate hybrid precoders presents two significant challenges. First, accurate Channel State Information (CSI) is essential for effective precoding. Obtaining precise CSI in a high dimensional hybrid system is a complex task in itself \cite{alkhateeb2014}. Second, even with accurate CSI, solving the optimization problem to design hybrid precoders is computationally challenging \cite{sohrabi2016}. These difficulties are further exacerbated by hardware impairments, such as an imperfect antenna array, which can cause classical model-based methods to fail.

Addressing these challenges requires innovative approaches that go beyond traditional methods. Purely data-driven methods, such as deep learning, have shown potential in handling complex wireless communication problems \cite{oshea2017}. However, these methods often require vast amounts of data and computational resources, making them impractical for real-time applications. Model-based learning \cite{shlezinger2023}, which combines the strengths of both model-based and data-driven approaches, offers a more balanced solution. By incorporating domain knowledge into the learning process, model-based learning has the potential to achieve better performance with reduced computational complexity.

Both channel estimation \cite{yassine2022,chatelier2023,mateosramos2025} and hybrid precoding \cite{lavi2023,nguyen2023,shlezinger2024} have been extensively studied using model-based learning approaches. However, these tasks have typically been tackled separately, without considering joint optimization that could offer potential advantages. 

\noindent \textbf{Contributions.} This paper proposes a novel framework that addresses channel estimation for a hybrid system and hybrid precoding \emph{jointly} using model-based learning techniques. By doing so, we aim to improve the overall performance and efficiency of hybrid massive MIMO systems, even in the presence of hardware impairments. In more details, the main contributions of this paper are:
\begin{itemize}[leftmargin=*]
    \item A deep unfolded neural network structure that jointly realizes channel estimation and hybrid precoding. This structure takes as inputs pilot-based measurements, to estimate the channel and tune the precoders. In contrast, the work in \cite{yassine2022} also begins with pilot-based measurements but only estimates the channel and does not consider hybrid systems, while \cite{lavi2023} assumes channels are available as inputs for tuning the precoders without considering channel estimation.
    \item Several training strategies of the proposed structure, depending on the availability of a clean CSI database.
    including:
i) \emph{Layer-by-layer} training, where the two tasks are trained separately using distinct cost functions. This approach can be implemented in either a supervised manner 
or an unsupervised one;
ii) \emph{End-to-end} training, where the entire network is trained jointly to maximize the sum-rate. This method requires supervised learning. 
    \item An extensive set of experiments to validate the proposed methods and highlight their applicability in practical scenarios.
\end{itemize}
Overall, the proposed method relies on only a few dozen learnable parameters while still achieving performance comparable to that of precoder computation based on perfect channel estimates, even when trained in an unsupervised manner.


\section{System model}

\begin{figure*}[t]
\centering
\includegraphics[width=0.7\textwidth]{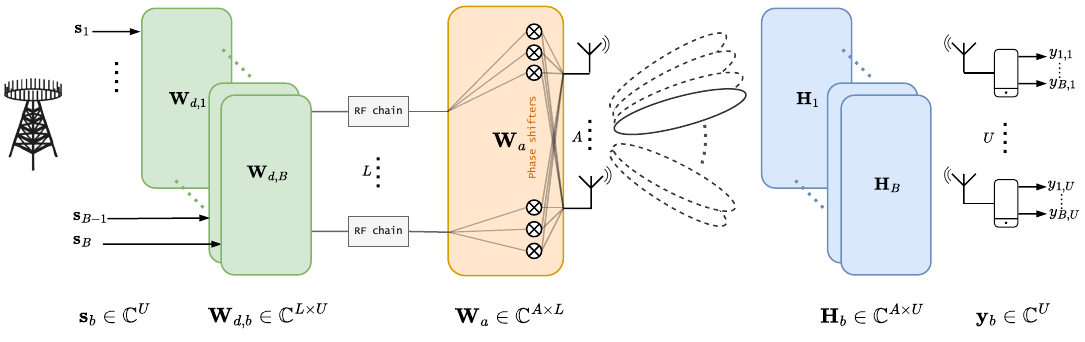}
\caption{Architecture of the considered hybrid MIMO system}
\label{fig:hybrid_MIMO_archi}
\end{figure*}

In this paper, a wideband hybrid multiuser-MIMO (MU-MIMO) system is considered. The Base Station (BS), equipped with \( A \) antennas, communicates with \( U \) single-antenna users and operates across \( B \) frequency subbands, which are shared by all users in a non-orthogonal manner. The BS employs hybrid beamforming, which is a compromise between fully digital and analog beamforming, where the number of RF chains \( L \) is limited (\( L < A \)). 
The digital precoder is represented by \( \mathbf{W}_{d,b} \in \mathbb{C}^{L \times U} \) for each frequency subband $b \in B$, while the analog precoder is represented by \( \mathbf{W}_a \in \mathbb{C}^{A \times L} \) (it is the same for all subbands).
The architecture of the system is illustrated in Figure \ref{fig:hybrid_MIMO_archi}. To simplify notation, we consider a single subcarrier, omitting the frequency bin index $b$ throughout the remainder of this paper.

\subsection{Observation model}\label{observ_model}

The method proposed aims to jointly perform the two tasks of uplink channel estimation and hybrid downlink precoding optimization based on received uplink pilots.

\subsubsection{Uplink model}
In this phase, users are assumed to transmit mutually orthogonal pilot signals.
The BS receives the uplink signals at the RF chain level, with analog precoding applied through \( \mathbf{W}_a \). After correlating with the user's pilot sequence, the resulting signal at time \( t \) 
is expressed as \( \mathbf{W}_a^{\mathsf{H}}(t) (\mathbf{h}_u + \mathbf{n}_{\text{UL},u}(t)) \in \mathbb{C}^{L} \), where $\mathbf{h}_u \in \mathbb{C}^A$ is the channel and $\mathbf{n}_{\text{UL},u}(t) \sim  \mathcal{CN}(0, \zeta^2 \mathbf{I}_A)$ is the received noise for user $u$.

Assuming that channel estimation is performed based on measurements collected over \( T \) time frames, we denote $\mathbf{M} \triangleq \begin{bmatrix} \mathbf{W_a}^* (1), \dots, \mathbf{W_a}^* (T) \end{bmatrix}^{\top} \in \mathbb{C}^{TL \times A}$ as the concatenation of the analog combiners used during the observation phase. 
Similarly, we denote $\mathbf{n}_{\text{UL},u} \triangleq \begin{bmatrix} \mathbf{n}_u(1)^\top, \dots, \mathbf{n}_u(T)^\top \end{bmatrix}^\top \in \mathbb{C}^{AT}$ as the concatenation of the received noise over the observation phase.

To model the structured impact of the analog combiners on the noise, we define  
\[\small
\tilde{\mathbf{M}} \triangleq
\begin{bmatrix}
\mathbf{W}_a^{\mathsf{H}} (1) &  \cdots & \mathbf{0} \\
\vdots  & \ddots & \vdots \\
\mathbf{0} &  \cdots & \mathbf{W}_a^{\mathsf{H}} (T)
\end{bmatrix} \in \mathbb{C}^{TL \times AT}.
\]  
With this definition, the received signal for the \( u \)-th user is given by  
$
\label{eq:ul_obs_u}
\mathbf{y}_{\text{UL},u} = \mathbf{M} \mathbf{h}_u + \tilde{\mathbf{n}}_{\text{UL},u}\in \mathbb{C}^{TL},$ where $ \tilde{\mathbf{n}}_{\text{UL},u} \triangleq \tilde{\mathbf{M}}\mathbf{n}_{\text{UL},u}.$
Forming a matrix from the signals of all users, we obtain:  
\begin{equation}
\label{eq:ul_obs}
\mathbf{Y}_{\text{UL}} = \mathbf{M} \mathbf{H} + \tilde{\mathbf{N}}_{\text{UL}} \in \mathbb{C}^{TL \times U}, 
\end{equation}
where \(\mathbf{H} \triangleq (\mathbf{h}_1, \dots, \mathbf{h}_U) \in \mathbb{C}^{A \times U}\) represents the channel, and \(\tilde{\mathbf{N}}_{\text{UL}} \triangleq (\tilde{\mathbf{n}}_{\text{UL},1}, \dots, \tilde{\mathbf{n}}_{\text{UL},U}) \in \mathbb{C}^{TL \times U}\) is the noise matrix. The average uplink signal-to-noise ratio (SNR) is given by  
$
\text{SNR}_{\text{av}, \text{UL}} = \frac{1}{U} \sum\nolimits_{u=1}^{U} \frac{\left\| \mathbf{h}_u \right\|_2^2}{A \cdot \zeta^2}.
$
Consequently, the channel estimation problem involves estimating the uplink channel \(\mathbf{H}\) based on the received signal \(\mathbf{Y}_{\text{UL}}\) and the measurement matrix \(\mathbf{M}\). The matrix comprising the channel estimates for all \(U\) users is denoted as \(\hat{\mathbf{H}} \in \mathbb{C}^{A \times U}\).

\subsubsection{Downlink model}
The downlink signal received by the $u$-th user is expressed as $y_{\text{DL},u} = \mathbf{h}_u^\top \mathbf{W} \mathbf{s} + n_{\text{DL},u}$, where $\mathbf{s} \in \mathbb{C}^U$ is the transmitted signal vector such that $\mathbb{E}[\mathbf{s}\mathbf{s}^{\mathsf{H}}]=\frac{1}{U}\mathbf{I}_U$, $n_{\text{DL},u}\sim  \mathcal{CN}(0, \sigma^2)$ is noise and $\mathbf{W} = \mathbf{W}_a\mathbf{W}_d$. 
Once again, concatenating received signals by all users yields
\begin{equation}
    \mathbf{y}_{\text{DL}} = \mathbf{H}^\top \mathbf{W} \mathbf{s} + \mathbf{n}_{\text{DL}} \in \mathbb{C}^{U},
\end{equation}
 with $\mathbf{y}_{\text{DL}}\triangleq(y_{\text{DL},1},\dots,y_{\text{DL},U})^\top$ and the noise $\mathbf{n}_{\text{DL}}\triangleq(n_{\text{DL},1},\dots,n_{\text{DL},U})^\top \sim \mathcal{CN}(0, \sigma^2 \mathbf{I}_U)$. The average downlink SNR is given by
$ \text{SNR}_{\text{av},\text{DL}}=\frac{1}{U}\sum_{u=1}^U{\frac{\left\| \mathbf{h}_u \right\|_2^2}{\sigma^2}}$.

\subsection{System constraints}
 

The proposed hybrid system incorporates two constraints:
\subsubsection{Total power constraint} 
The total downlink power is limited as
$
    \| \mathbf{W}_a \mathbf{W}_d \|_F^2 \leq P_{\text{total}},
$
where \( \| \cdot \|_F \) denotes the Frobenius norm.

\subsubsection{Hardware constraint - Phase shifter networks}
We focus on a typical phase-shifter-based architecture from \cite{Mendez-Rial2016}, where each transceiver connects to antennas via phase shifters. This enforces a unit-modulus constraint on the entries of \( \mathbf{W}_a \), such that \( \mathbf{W}_a \in \mathcal{G} \), where  
$
    \mathcal{G} \triangleq \left\{ \mathbf{A}  \in \mathbb{C}^{A \times L} \mid | a_{ij}| = 1, \ \forall (i,j) \right\}
$.

\subsection{Channel model}
 Let \( \{ \overrightarrow{a_1}, \dots, \overrightarrow{a_A} \} \) denote the locations of the antennas (assumed isotropic) relative to the centroid of the antenna array. Let \( \lambda \) be the wavelength under consideration. Under the plane wave assumption, the channel resulting from a single propagation path with a direction of arrival (DoA) \( \overrightarrow{u} \) is proportional to the steering vector \( \mathbf{e}(\overrightarrow{u}) \), which is expressed as 
$
\mathbf{e}(\overrightarrow{u}) \triangleq \frac{1}{\sqrt{A}} \begin{pmatrix} 
 \mathrm{e}^{-\mathrm{j} \frac{2\pi}{\lambda} \overrightarrow{a_1} \cdot \overrightarrow{u}}, \dots,  \mathrm{e}^{-\mathrm{j} \frac{2\pi}{\lambda} \overrightarrow{a_A} \cdot \overrightarrow{u}} 
\end{pmatrix}^\top \in \mathbb{C}^A
$.
The multipath channel can then be expressed as a linear combination of steering vectors
$
\label{linear_combi}
    \mathbf{h} = \sum\nolimits_{p=1}^{P} \beta_p \mathbf{e}(\overrightarrow{u_p})$
where \( P \) is the number of propagation paths. 

Building a dictionary of steering vectors corresponding to \( N \) potential DoAs, denoted as
$
\mathbf{D} \triangleq \begin{pmatrix}
\mathbf{e}(\overrightarrow{u_1}), \dots, \mathbf{e}(\overrightarrow{u_N})
\end{pmatrix} \in \mathbb{C}^{A \times N},
$
 the channel \( \mathbf{h}_u \) for each user can be approximated as $ \mathbf{h}_u \approx \mathbf{D} \boldsymbol{\beta}_u,$
where the sparsity of \( \boldsymbol{\beta}_u \) is given by \( \|\boldsymbol{\beta}_u\|_0 = P \) corresponds to the number of propagation paths. Within this model, estimating the channel amounts to solve a sparse recovery problem in a dictionary of steering vectors. 
This can be tackled for example by greedy sparse recovery algorithms, such as matching pursuit (MP) \cite{mallat1993}.

\noindent \textbf{Hardware impairments in the BS system.}
The aforementioned approach assumes perfect knowledge of the dictionary. However, in real-world scenarios, very precise knowledge of antenna locations is often out of reach. To address this, mpNet \cite{yassine2022} is an unfolded neural network which proposes to learn the dictionary, initializing with an imperfect one referred to as the \textit{nominal dictionary} $\tilde{\mathbf{D}} \triangleq \begin{pmatrix}
\tilde{\mathbf{e}}(\overrightarrow{u_1}), \dots, \tilde{\mathbf{e}}(\overrightarrow{u_N}) \end{pmatrix}$, where $
\tilde{\mathbf{e}}(\overrightarrow{u}) \triangleq \frac{1}{\sqrt{A}} \begin{pmatrix} 
 \mathrm{e}^{-\mathrm{j} \frac{2\pi}{\lambda} \overrightarrow{\tilde{a}_1} \cdot \overrightarrow{u}}, \dots,  \mathrm{e}^{-\mathrm{j} \frac{2\pi}{\lambda} \overrightarrow{\tilde{a}_A} \cdot \overrightarrow{u}} 
\end{pmatrix}^\top
$, $\overrightarrow{\tilde{a}_i}$ being the nominal location of the $i$th antenna. 
This strategy aims to converge toward the ideal dictionary $\mathbf{D}$, which is referred to as the \textit{real dictionary}.

\section{Problem formulation} 
This work aims to optimize hybrid precoders while relying on channel estimates obtained from uplink pilots. Let us introduce the channel estimator $\hat{\mathbf{H}}_{\boldsymbol{\theta}}(\mathbf{Y}_{\text{UL}})$ (denoted $\hat{\mathbf{H}}_{\boldsymbol{\theta}}$ in the sequel for brevity reasons), the vector $\boldsymbol{\theta}$ gathering its parameters. In a similar way, let us introduce the precoding function $\mathbf{W}_{\boldsymbol{\mu}}(\hat{\mathbf{H}}_{\boldsymbol{\theta}})$, the vector $\boldsymbol{\mu}$ gathering its parameters. 

The goal is to maximize the achievable sum-rate. The corresponding optimization problem can be formulated in two ways, depending on whether a database of true channels is available or not. In the positive case, the problem is:

\begin{equation} \label{eq:supervised_problem}
\small
\begin{aligned}
    \underset{\boldsymbol{\mu}, \boldsymbol{\theta}}{\text{minimize}} \,\,\, & \mathcal{C}_\text{2,S}=-\log \det \left( \mathbf{I}_U + \tfrac{1}{U \sigma^2} \mathbf{H} \mathbf{W}_{\boldsymbol{\mu}}(\hat{\mathbf{H}}_{\boldsymbol{\theta}}) \mathbf{W}_{\boldsymbol{\mu}}^{\mathsf{H}}(\hat{\mathbf{H}}_{\boldsymbol{\theta}}) \mathbf{H}^{\mathsf{H}} \right) \\ &
    \text{where} \quad \mathbf{W}_{\boldsymbol{\mu}}(\hat{\mathbf{H}}_{\boldsymbol{\theta}}) = \mathbf{W}_{a,{\boldsymbol{\mu}}}(\hat{\mathbf{H}}_{\boldsymbol{\theta}}) \mathbf{W}_{d,{\boldsymbol{\mu}}}(\hat{\mathbf{H}}_{\boldsymbol{\theta}}), \\
    \text{subject to} \quad & \| \mathbf{W}_{\boldsymbol{\mu}}(\hat{\mathbf{H}}_{\boldsymbol{\theta}}) \|_F^2 \leq P_{\text{total}},\mathbf{W}_{a,{\boldsymbol{\mu}}}(\hat{\mathbf{H}}_{\boldsymbol{\theta}}) \in \mathcal{G}. 
\end{aligned}
\end{equation}
Here, the subscript S stands for supervised learning.
This objective can be interpreted as maximizing the achievable sum-rate based on channel estimates. Note that the problem tackled in \cite{lavi2023} is similar, except that perfect channels are given as inputs to the precoding function ($\mathbf{W}_{\boldsymbol{\mu}}(\mathbf{H})$ instead of $\mathbf{W}_{\boldsymbol{\mu}}(\hat{\mathbf{H}}_{\boldsymbol{\theta}})$), which is the only one being optimized (channel estimation is not considered). 
This problem can be tackled either as is, or sequentially by first optimizing the channel estimator as follows:
\begin{equation}\label{eq:Cost_funct_mpnet_sup}
\small
    \underset{\boldsymbol{\theta}}{\text{minimize}} \,\,\, \mathcal{C}_\text{1,S} = \frac{\|\hat{\mathbf{H}}_{\boldsymbol{\theta}} - \mathbf{H}\|_F^2}{\|\mathbf{H}\|_F^2}
\end{equation}
and then injecting the obtained parameter value $\boldsymbol{\theta}^\star$ into the original problem \eqref{eq:supervised_problem}.

Alternatively, if a database of true channels is not available, the problem becomes:
\begin{equation} \label{eq:unsupervised_problem}
\small
\begin{aligned}
    \underset{\boldsymbol{\mu}}{\text{minimize}} \,\,\, & \mathcal{C}_\text{2,U}=-\log \det \left( \mathbf{I}_U + \tfrac{1}{U \sigma^2} \hat{\mathbf{H}}_{\boldsymbol{\theta}} \mathbf{W}_{\boldsymbol{\mu}}(\hat{\mathbf{H}}_{\boldsymbol{\theta}}) \mathbf{W}_{\boldsymbol{\mu}}^{\mathsf{H}}(\hat{\mathbf{H}}_{\boldsymbol{\theta}}) \hat{\mathbf{H}}_{\boldsymbol{\theta}}^{\mathsf{H}} \right) \\
    \text{subject to} \,\,\, & \| \mathbf{W}_{\boldsymbol{\mu}}(\hat{\mathbf{H}}_{\boldsymbol{\theta}}) \|_F^2 \leq P_{\text{total}},\mathbf{W}_{a,{\boldsymbol{\mu}}}(\hat{\mathbf{H}}_{\boldsymbol{\theta}}) \in \mathcal{G}. 
\end{aligned}
\end{equation}
The subscript U refers to the unsupervised learning case, where the objective can be interpreted as maximizing an estimate of the achievable sum-rate based on channel estimates.
This problem differs from \eqref{eq:supervised_problem} in two ways: the true channels in the sum-rate formula are replaced by their estimation, and, as a consequence, optimization is done only on the precoding parameters $\boldsymbol{\mu}$. Indeed, jointly optimizing $\boldsymbol{\mu}$ and $\boldsymbol{\theta}$ would be ill-posed in this context, as the cost function would not depend on the observation matrix $\mathbf{Y}_\mathrm{UL}$. 
Then, the value of $\boldsymbol{\theta}$ to be used within \eqref{eq:unsupervised_problem} necessarily comes from an optimization of the channel estimator as follows: 
\begin{equation} \label{eq:Cost_funct_mpnet_unsup}
\small
    \underset{\boldsymbol{\theta}}{\text{minimize}} \,\,\, \mathcal{C}_\text{1,U} = \frac{\|\mathbf{M}\hat{\mathbf{H}}_{\boldsymbol{\theta}} - \mathbf{Y}_{\text{UL}}\|_F^2}{\|\mathbf{Y}_{\text{UL}}\|_F^2}.
\end{equation}
Note that the channel is reconstructed in \eqref{eq:Cost_funct_mpnet_sup} while the observation is reconstructed in \eqref{eq:Cost_funct_mpnet_unsup} (since a clean channel database is not available).

\section{Proposed approach}
\begin{figure}[h]
\centering
\includegraphics[width=\columnwidth]{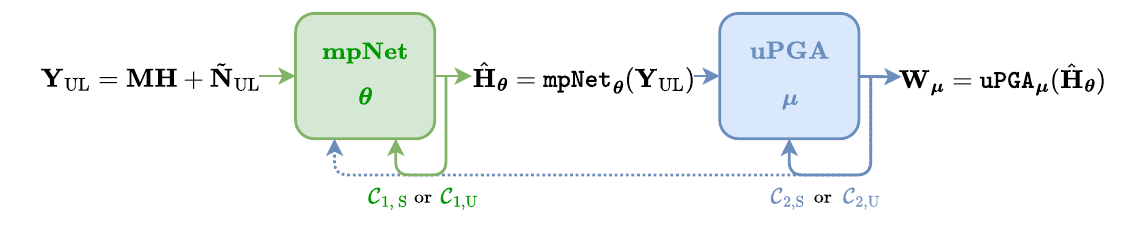}
\caption{Schematic of the proposed approach}
\label{fig:proposed_structure}
\vspace{-0.2cm}
\end{figure}

This paper proposes to use deep unfolding of iterative algorithms for both the channel estimator $\hat{\mathbf{H}}_{\boldsymbol{\theta}}$ and the precoding function $\mathbf{W}_{\boldsymbol{\mu}}$. Indeed, the proposed structure combines unfolded matching pursuit (mpNet) \cite{yassine2022} for channel estimation and unfolded projected gradient ascent (unfolded PGA) \cite{lavi2023} for precoder design. The complete architecture is presented in Figure \ref{fig:proposed_structure}. 


\begin{itemize}[leftmargin=*]
    \item \emph{mpNet-Based Channel Estimation:}\label{mpNet_explanation} Starting from the uplink observation $\mathbf{Y}_{\text{UL}}$ of \eqref{eq:ul_obs}, the first step aims at obtaining a channel estimate $\hat{\mathbf{h}}_u \in \mathbb{C}^{A}$ for each user $u \in \{1,\dots,U\}$. This can be done by feeding independently each column of $\mathbf{Y}_{\text{UL}}$ to mpNet, which is an unfolded neural network performing sparse recovery with the matching pursuit algorithm using a dictionary of parameterized steering vectors. Finally, forming a matrix from the estimated channels $\hat{\mathbf{h}}_u \in \mathbb{C}^{A},u \in \{1,\dots,U\}$ as columns yields 
    \[ \hat{\mathbf{H}}_{\boldsymbol{\theta}} = \texttt{mpNet}_{\boldsymbol{\theta}}(\mathbf{Y}_{\text{UL}}) \in \mathbb{C}^{A \times U},\]
    where $\boldsymbol{\theta}$ denotes the learnable parameters of mpNet.
    To account for hardware impairments,  the weight matrix is initialized using the nominal dictionary of steering vectors, denoted by $\tilde{\mathbf{D}}$, which is progressively refined during training to converge toward the true dictionary. Two cases are considered for the training of the dictionary: either i) each entry of the dictionary is a parameter in which case $\boldsymbol{\theta} \in \mathbb{C}^{AN}$ (i.e., $2AN$ real parameters) and the method is called \emph{unconstrained} mpNet \cite{yassine2022}, or ii) only the physical parameters defining the steering vectors (antenna locations) are parameters, in which case $\boldsymbol{\theta} \in \mathbb{R}^{A}$(i.e., $A$ real parameters) and the method is called \emph{constrained} mpNet \cite{chatelier2023}.
    
    \item \emph{Unfolded PGA-Based Precoding Optimization:} The second step uses the previously obtained channel estimates $\hat{\mathbf{H}}$ to optimize precoding. To do so, an unfolded version of the projected gradient ascent algorithm with the sum-rate as objective function is used. The resulting precoder is of the form
    \[
    \mathbf{W}_{\boldsymbol{\mu}}= \texttt{uPGA}_{\boldsymbol{\mu}}(\hat{\mathbf{H}}_{\boldsymbol{\theta}})= \texttt{uPGA}_{\boldsymbol{\mu}}(\texttt{mpNet}_{\boldsymbol{\theta}}(\mathbf{Y}_{\text{UL}})),
    \]
    where $\boldsymbol{\mu}$ denotes the learnable parameters of the unfolded PGA algorithm, which correspond to the step sizes used during the iterative updates of the analog and digital precoders. For \( K \) iterations of the unfolded algorithm, we define \( \boldsymbol{\mu} \in \mathbb{R}^{K \times 2} \), where the step size at the \( k \)-th iteration is given by \( \boldsymbol{\mu}_k \triangleq [\mu_a^{(k)}, \mu_d^{(k)}] \), with \( \mu_a^{(k)} \) and \( \mu_d^{(k)} \) representing the step sizes for the analog and digital precoders, respectively. The complete step size matrix is expressed as $\boldsymbol{\mu} = \begin{pmatrix}
        \boldsymbol{\mu}_0, \dots, \boldsymbol{\mu}_{K-1}
    \end{pmatrix}^\top.$
    Initially, the step sizes \( \boldsymbol{\mu} \) are set using a fixed hyperparameter and are subsequently optimized throughout training to maximize the achievable sum-rate. As in \cite{lavi2023}, the starting point for the gradient ascent of the analog precoding matrix \( \mathbf{W}_a^{(0)} \) is obtained using the first \( L \) right-singular vectors of the estimated channel \( \hat{\mathbf{H}}_{\boldsymbol{\theta}} \), with unit-modulus constraints enforced by normalizing the entries. As for the digital precoder \( \mathbf{W}_d^{(0)} \), it is initialized randomly.

\end{itemize}

\noindent \textbf{Training strategy.}
The previous paragraph presented two optimization problems: one corresponding to supervised learning in \eqref{eq:supervised_problem}, and the other to unsupervised learning in \eqref{eq:unsupervised_problem}. This paragraph outlines the training strategies associated with each problem.

\begin{itemize}[leftmargin=*]
    \item \emph{Supervised Learning}
    \begin{enumerate}[label=\arabic*), leftmargin=*, start=1]
        \item \textbf{End-to-End cold start:} This strategy involves directly combining the two unfolded models and training the entire architecture in an end-to-end manner, with the objective of minimizing the loss function $\mathcal{C}_{2,\text{S}}$ defined in~\eqref{eq:supervised_problem}.
        
        \item \textbf{Layer-by-Layer training:} This approach decomposes the training process into two stages. First, the channel estimation model, mpNet, is trained by minimizing $\mathcal{C}_{1,\text{S}}$ as defined in~\eqref{eq:Cost_funct_mpnet_sup}. The resulting channel estimate is then used to train the precoding model, unfolded PGA, by minimizing $\mathcal{C}_{2,\text{S}}$.
        
        \item \textbf{End-to-End warm start:} This strategy extends the previous one by introducing a final end-to-end training phase. After the two components have been trained separately, the entire system is fine-tuned jointly to minimize $\mathcal{C}_{2,\text{S}}$.
    \end{enumerate}
    
    \item \emph{Unsupervised Learning}
    \begin{enumerate}[label=\arabic*), leftmargin=*, resume]
        \item \textbf{Layer-by-Layer training:} Analogous to the supervised learning case, this strategy first trains the channel estimation model by minimizing $\mathcal{C}_{1,\text{U}}$ as defined in~\eqref{eq:Cost_funct_mpnet_unsup}. The resulting channel estimates are then used to train the precoding model by minimizing the loss $\mathcal{C}_{2,\text{U}}$ in~\eqref{eq:unsupervised_problem}.
    \end{enumerate}
\end{itemize}


\section{Experimental validation}
All experiments are performed using a dataset consisting of 30,000 training channels and 1,000 test channels at a frequency of $28$~GHz generated from Sionna RT \cite{sionna}, a realistic synthetic channel generator, in an urban environment. The antenna array under consideration is a uniform linear array (ULA) with $A = 64$ antennas, comprising impairments taking the form of uncertainties about antenna locations.  The nominal antenna locations $\overrightarrow{\tilde{a}_i}$ define a $\lambda/2$-separated ULA, and the true antenna locations are not evenly spaced and defined as
\vspace{-0.1cm}
\[
\vec{a}_i = \overrightarrow{\tilde{a}_i} + \lambda \mathbf{n}_{p,i}, \quad \mathbf{n}_{p,i} = (\eta_{p,i}, 0, 0)^\top, \quad \eta_{p,i} \sim \mathcal{N}(0, 10^{-2}).
\vspace{-0.1cm}
\] 
The uplink ad downlink noise variances are equal, $\sigma^2=\zeta^2$, and $\text{SNR}_{\text{av,UL}}$ values of $5$ and $15$ dB are tested.
The system comprises $L=16$ RF chains and the number of time frames $T$ varies between $1$ and $3$.
It is worth noting that the previous version of mpNet \cite{yassine2022} was also evaluated under a planar array configuration and in scenarios involving dynamic hardware impairments.
To gather the observation required for channel estimation \eqref{eq:ul_obs}, we consider an agnostic measurement matrix $\mathbf{M} \in \mathbb{C}^{TL \times A}$, defined as $ m_{ij} = \mathrm{e}^{\mathrm{j} \phi_{ij}}, \quad \phi_{ij} \sim \mathcal{U}[0, 2\pi]. $

\subsection{Unfolded Channel estimation for hybrid MIMO}
The first step towards the proposed joint estimation/precoding method is the extension of mpNet to hybrid MIMO systems and supervised training. Indeed, the previous versions of mpNet \cite{yassine2022,chatelier2023} were only designed for fully digital MIMO systems and trained in an unsupervised manner. To this aim, four variations of mpNet are assessed, incorporating supervised or unsupervised learning approaches, and either imposing or omitting dictionary constraints, as detailed in Section \ref{mpNet_explanation}.  

The performance of mpNet is compared to several baselines, including iterative MP initialized with both the real (unknown) dictionary and the nominal dictionary, which represents incomplete system knowledge. Additionally, we compare mpNet against a model-agnostic oracle linear minimum mean square error (LMMSE) estimator having access to the channel entries power. The details of the implementation are provided in a public repository\footnote{\href{https://github.com/nklaimi01/MoBAIWL}{See: https://github.com/nklaimi01/MoBAIWL}}. Three scenarios are considered, varying either the average SNR or the number of measurements, which depend on the number of RF chains $L$ and the number of time frames $T$. 


\begin{figure*}[t]
    \centering
    \includegraphics[width=\textwidth]{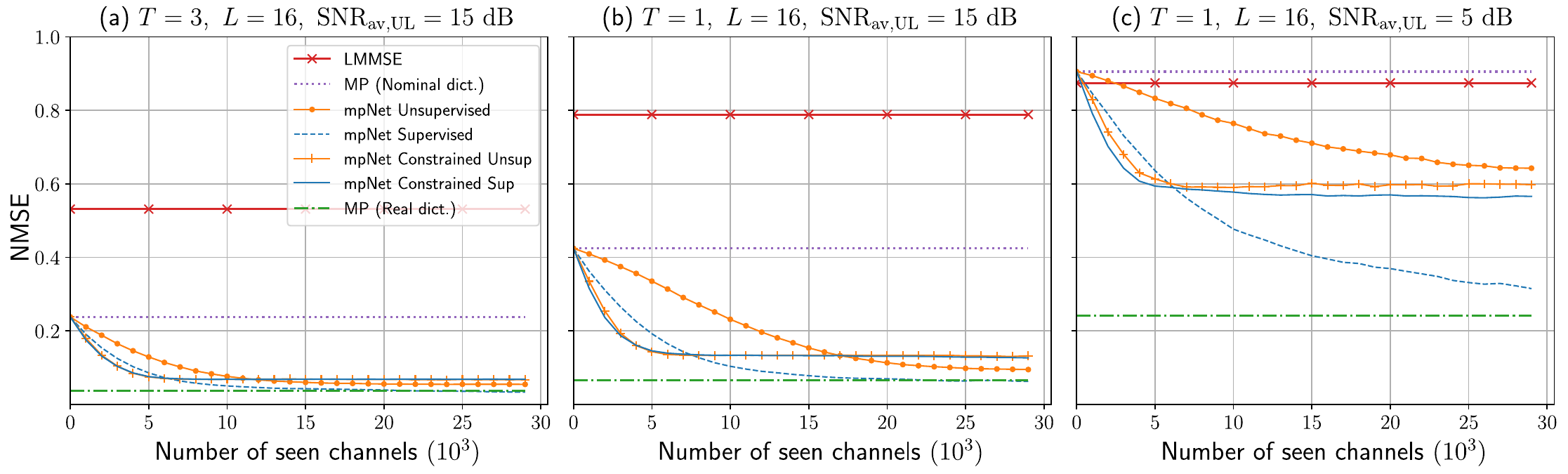}
    \captionsetup{font=normalsize}
    \caption{Channel estimation performance on unseen channels for different average SNR levels and numbers of time frames.}
    \label{fig:nmse_subfigures}
    ~\vspace{-0.85cm}
\end{figure*}

The results are shown in Figure \ref{fig:nmse_subfigures} and demonstrate that the mpNet methods outperform both LMMSE and MP with a nominal dictionary (which is the initialization of all mpNet versions). Furthermore, mpNet successfully converges toward the true system knowledge (except at low SNR with a few observations), achieving performance comparable to MP with the real dictionary and showing signs of surpassing it under high SNR conditions.
As expected, all methods perform better if the SNR is increased or if the number of observations $LT$ grows.

In unsupervised learning, mpNet with a constrained dictionary converges significantly faster than its unconstrained counterpart. However, in supervised learning, while the constrained approach achieves faster convergence, the unconstrained version ultimately achieves superior performance, albeit at a slower rate. This is because the unconstrained supervised learning approach allows the model to move beyond the limitations imposed by the physical constraints of the dictionary.  

In conclusion, mpNet exhibits a qualitatively similar behavior in hybrid MIMO systems to that of fully digital MIMO systems, \cite{yassine2022,chatelier2023}, and can be trained in a supervised way. Moreover, based on the findings described in the previous paragraphs, we adopt in subsequent experiments the constrained mpNet model for unsupervised learning and the unconstrained mpNet model for supervised learning.  

\subsection{Joint channel estimation and hybrid precoding}
This section evaluates the effectiveness of the proposed approach in terms of sum-rate. The proposed methods are compared against the achievable sum-rate in a fully digital system, which serves as an upper bound (i.e., the best possible performance). Additionally, they are compared to uPGA with full knowledge of the true channel (as in \cite{lavi2023}), as well as uPGA using LMMSE-based channel estimations as input, which represents a baseline.

\begin{figure}[h]
    \centering
    \includegraphics[width=0.4\textwidth]{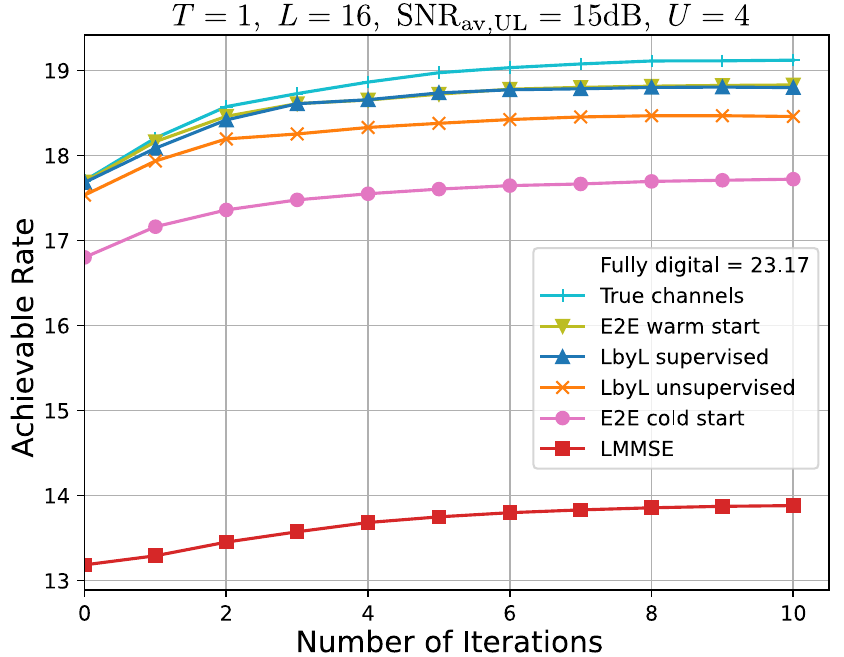}
    \captionsetup{font=normalsize}
    \caption{Sum-rate per PGA iteration.}
    \label{fig:sumrate_vs_iterations}
    ~\vspace{-0.65cm}
\end{figure}

The sum-rate results are shown in Figure \ref{fig:sumrate_vs_iterations} as a function of the number on unfolded uPGA iterations. As expected, the method taking true channels as inputs is best while the one using LMMSE channel estimation (model-free) is the worst.

First, we test an end-to-end training of the joint model with random initialization. As shown in Figure \ref{fig:sumrate_vs_iterations} (labeled ``E2E cold start"), this direct approach yields much better results than LMMSE thanks to the model-based estimation it performs, but is still quite far from the performance obtained using true channels.  

Next, we adopt a layer-by-layer training strategy, in which mpNet is first trained independently (Figure \ref{fig:nmse_subfigures}b), and then its channel estimates are used to train uPGA. This approach is evaluated in both supervised  and unsupervised  settings. The results show a significant improvement in sum-rate compared to the direct one, approaching the performance obtained with true channel information. 
Remarkably, the unsupervised approach yields sum-rates very close to the supervised one, despite the fact that it does not require a database of clean channels, and is thus much more practical to apply in real-world scenarios.

Finally, a fine-tuning step is applied by training end-to-end the joint model while initializing the model using the parameters obtained from the layer-by-layer supervised training (referred to as ``E2E warm start"). However, this step yields only marginal improvements while significantly increasing computational complexity, indicating that the conducted pretraining was sufficient. 
It is important to note that the unsupervised approach, since it uses a constrained dictionary, has significantly fewer parameters to learn (i.e., \( A + 2K = 84 \) parameters) compared to the three supervised approaches tested (i.e., \( 2AN + 2K = 153{,}620 \) parameters).

\section{Conclusion and Perspectives}

In this paper, we introduced a model for tuning hybrid precoders without requiring knowledge of the true channel state information. Experimental results demonstrate that, starting from measured pilot signals, the proposed model can accurately estimate the channel and leverage these estimates to optimize the hybrid precoders, achieving sum-rate performance close to that obtained with perfect channel state information.

The strong performance of the proposed unsupervised learning approach is a promising result, as it suggests that hybrid precoding can be effectively optimized in real-world scenarios where building a true channels database is difficult.


\bibliographystyle{IEEEtran} 
\bibliography{references} 

\end{document}